\documentclass[manuscript,screen,review=false]{acmart}
\usepackage{tikz}
\usepackage{caption} 
\usepackage{algpseudocode}
\usepackage{algorithm}
\usepackage{amsmath}
\usetikzlibrary{automata}
\usepackage{enumerate}
\usepackage{titlesec}
\usepackage{pgfplots}
\usepackage{multirow}
\usepackage{comment}
\pgfplotsset{width=8cm, height = 5cm}
\usepackage{graphicx}
\usepackage{comment}
\usepackage{caption}
\usepackage{subcaption}

\usepackage[utf8]{inputenc}
\usepackage{tabularx,booktabs}
\newcolumntype{C}{>{\centering\arraybackslash}X} 
\usepackage{pgfplots}
\usepackage{multirow}
\usepackage{adjustbox}
\pgfplotsset{width=10cm,compat=1.9}
\usepackage{tikz}
\usetikzlibrary{shapes.geometric, arrows}
\tikzstyle{line}= [draw, -latex']

\tikzstyle{startstop} = [rectangle, rounded corners, 
minimum width=10cm, 
minimum height=2cm,
text centered, text width=10cm,
draw=black]

\tikzstyle{io} = [trapezium, 
trapezium stretches=true, 
trapezium left angle=70, 
trapezium right angle=110, 
minimum width=3cm, 
minimum height=1cm, text centered, 
draw=black]

\tikzstyle{process} = [rectangle,  rounded corners,
minimum width=3cm, 
minimum height=1cm, 
text centered, text width=10cm,
draw=black]

\tikzstyle{decision} = [diamond, 
minimum width=10cm, 
minimum height=1cm, 
text centered, 
draw=black]
\tikzstyle{arrow} = [thick,->,>=stealth]

\AtBeginDocument{%
  \providecommand\BibTeX{{%
    \normalfont B\kern-0.5em{\scshape i\kern-0.25em b}\kern-0.8em\TeX}}}

\setcopyright{acmcopyright}
\copyrightyear{2018}
\acmYear{2018}
\acmDOI{XXXXXXX.XXXXXXX}

\begin{document}

\title{GPU Acceleration of a Conjugate Exponential
Model for Cancer Tissue Heterogeneity}

\author{Anik Chaudhuri}
\email{anik.chaudhuri09@gmail.com}
\affiliation{%
  \institution{Indian Institute of Technology Bhubaneswar}
  \city{Bhubaneswar}
  \state{Odisha}
  \country{India}
}

\author{Anwoy Kumar Mohanty}
\affiliation{%
  \institution{The Arena Group}
  \country{New York}
  }
\email{anwoy.rkl@gmail.com}

\author{Manoranjan Satpathy}
\affiliation{%
  \institution{Indian Institute of Technology Bhubaneswar}
  \city{Bhubaneswar}
  \country{India}
}
\email{manoranjan@iitbbs.ac.in}

\renewcommand{\shortauthors}{Chaudhuri, et al.}

\begin{abstract}
  Heterogeneity in the cell population of cancer tissues poses many challenges in cancer diagnosis and treatment. Studying the heterogeneity in cell populations from gene expression measurement data in the context of cancer research is a problem of paramount importance. In addition, reducing the computation time of the algorithms that deal with high volumes of data has its obvious merits.  Parallelizable models using Markov chain Monte Carlo methods are typically slow. This paper shows a novel, computationally efficient, and parallelizable model to analyze heterogeneity in cancer tissues using GPUs. Because our model is parallelizable, the input data size does not affect the computation time much, provided the hardware resources are not exhausted.
Our model uses qPCR (quantitative polymerase chain reaction) gene expression measurements to study heterogeneity in cancer tissue. We compute the cell proportion breakup by accelerating variational methods on a GPU. We test this model on synthetic and real-world gene expression data collected from fibroblasts and compare the performance of our algorithm with those of MCMC and Expectation Maximization. Our new model is computationally less complex and faster than existing Bayesian models for cancer tissue heterogeneity.
\end{abstract}




\maketitle

\section{Introduction}
The clonal evolution model is a well-known model to explain proliferation in cancer progression \cite{a} and \cite{b}. There is a widespread belief that tumors arise from a single mutated cell as described in \cite{a} and \cite{b}; thereafter, accumulated mutation makes it neoplastic. According to the clonal evolution model, the cell population becomes heterogeneous due to the stepwise accumulation of mutation. The stem cell theory states that tumor cells have a small proportion of dominant cells out of the entire cell population \cite{c}, \cite{d}, \cite{e}, and \cite{f}. These theories explain the heterogeneous cell population in cancer tissues. Heterogeneity in a cell population gives rise to many challenges because therapies behave differently for sub-populations of the population. A particular treatment may be helpful for a specific cell population for a patient; however, the same treatment may not be a good fit for some other patient with a different arrangement of cell subpopulation. So, it is very important to model this heterogeneous behavior of cancer cells. A good treatment decision, like the amount of drug that needs to be given, can only be made if the breakup of the heterogeneous cell population is known beforehand.\par
Authors in \cite{28} presented a comparison of different Bayesian models for leukemia data. Magnetic resonance imaging (MRI) data was used to investigate tissue heterogeneity in prostate cancer in \cite{27}. In \cite{h}, a parallelizable model for cancer tissue heterogeneity was presented.
Hierarchical models were used in \cite{g} and \cite{8} to model heterogeneity in cancer tissues. Boolean networks to find out the proportion-wise breakup of each subpopulation were used in \cite{g} and \cite{8}. The interaction between the Boolean networks and the observed gene expression data was modeled by a multilevel hierarchical model. The same result was obtained by the authors in \cite{8} by using variational Bayes. However, these models can not reap the benefits of GPU parallelism because they are not suitable for the SIMD (Single Instruction Multiple Data) architecture of a GPU.\par
In \cite{h}, the authors demonstrated a parallelizable model to analyze heterogeneity in cancer tissues by using the Markov chain Monte Carlo (MCMC) algorithm. The authors in \cite{h} used the Metropolis$-$Hastings (M-H) algorithm on a GPU to calculate the posterior distribution of the weight of each subpopulation. The model used gene expression measurements from fibroblasts to compute the weights of each subpopulation. This method is compute-intensive, and it is difficult to judge convergence because it is not easy to know when the chain converges. It also requires additional thinning to obtain a good effective sample size, which increases computation time. In this paper, we overcome these problems by utilizing a new parallelizable model that allows us to use variational Bayes to estimate the posterior of the unknown parameters on a GPU; in short, our work is a continuation of the research in \cite{h}.

We highlight our main contributions in the following headings:\\
\textbf{Reduced computational complexity:}\newline
This paper discusses a new approach to analyzing cancer tissue heterogeneity which is computationally less complex than \cite{g} and \cite{8}. The models in \cite{g} and \cite{8} are not parallelizable, so their complexity increases with the input data size, so computation time increases with an increase in input data size. In contrast, our parallelizable model can finish computation much faster than \cite{g} and \cite{8} because the computation time does not depend on the data size as long as hardware resources are not exhausted.\\

\noindent \textbf{Resolves convergence issue of previous parallelizable model:}\newline
Our new parallelizable conjugate model converges much faster than \cite{h}. The model in \cite{h} takes a lot of iterations to converge, but our model converges in a significantly lesser number of iterations when compared to \cite{h}. Therefore, our new parallelizable model is superior in terms of reduced complexity and faster convergence when compared to the existing Bayesian models used to analyze cancer tissue heterogeneity. Section 3 presents the details.\\

\noindent \textbf{Comparison with state-of-the-art:}\newline
The proposed model improves state of the art model, i.e., \cite{h}. The authors in \cite{h} used a Markov chain Monte Carlo algorithm called Metropolis-Hastings, which has convergence issues. Our proposed model takes care of the convergence issue by using variational Bayes. This paper will also show the advantage of accelerating the variational Bayes and expectation-maximization algorithm on a GPU. We have compared the complexity of our model with the state-of-the-art and other Bayesian models used to analyze cancer tissue heterogeneity and show that our model converges quicker and reduces computation time. To check the correctness of our model we use synthetic data and experimental data to compute the unknown parameters. To demonstrate the superiority and correctness of our model, the experimental results are compared with the results of existing cancer tissue heterogeneity models. In addition, a multilevel hierarchical Bayesian model using variational Bayes and expectation maximization has not been accelerated for GPU yet; therefore, our research will be of interest to the research community in this field.\\
The paper is organized as follows. Section 2 discusses the necessary background and motivation for this research. Section 3 discusses our method, a comparison to current state-of-the-art models, where we discuss the complexity of our model. In section 4, we experiment with synthetic data; section 5 concludes the paper.

\section{Background and motivation}

Cellular interaction arises from the interaction between protein and DNA \cite{b} and \cite{9}. Cellular behavior has been modeled by gene regulatory network. Such networks have been modeled by using several methods, which include differential equation \cite{10}, deterministic and probabilistic Boolean network \cite{9}, \cite{11} and Bayesian and dynamic networks \cite{12} and \cite{13}. It is challenging to learn from real-world data due to the large search space of the parameters of the probabilistic Boolean network. Continuous valued measurements is used in the ARCANE method \cite{14} to find out regulatory interactions. The authors in \cite{15} used the Bayesian method to deduce pathway structures from quantitative genetic interaction observations. The REVEAL algorithm \cite{16} uses discretized time domain data to learn deterministic Boolean networks. A computational model of the Mycoplasm genitalium was described in \cite{17}. A good grasp of the different cell dynamics is needed to work with such a model, so this approach can deal only with simpler systems. Modeling cancer tissue is not suitable with such an approach .\par
The biological literature includes a great deal of research that discusses the interaction between biological molecules. This information used as prior knowledge can reduce the dependence on a large quantity of data. Cell behavior can be precisely modeled by such knowledge. The authors in \cite{18} took advantage of prior knowledge to create Boolean networks from marginal pathway knowledge by using Karnaugh maps. This method was used in \cite{19} to create a Boolean network of the Mitogen-Activated Protein Kinase (MAPK) signal transduction network. MAPK network is a widely studied network, and its Boolean network representation is shown in figure \ref{fig1}.
The Boolean network in figure \ref{fig1} represents the relation between various biological molecules and genes. A good number of reactions in the transduction network have binary states. For example, phosphorylated protein can be regarded as one and unphosphorylated protein can be regarded as zero. Therefore, a Boolean network with AND, OR, NOT logic gates can be used to represent their interaction exactly. \par
In \cite{19}, the authors represented cancer as a "stuck-at" fault. A variable being stuck in an on or off state permanently, irrespective of the upstream variables, is called a stuck-at-fault. For example, in figure \ref{fig1}, if EGFR is in \emph{on state} irrespective of EGF; this is stuck at one fault. An example of a stuck-at-zero fault is EGFR in an off state irrespective of EGF. This allows us to use the conventional digital electronics testing methods to be applied in analyzing signal transduction networks. This can then be used to design better cancer therapy.\par
The authors in \cite{19} considered only one stuck at fault at a time, but it may be extended to the case of numerous faults by only taking into account one stuck at fault at a time. In both cases, this method assumes that a single flawed network can adequately simulate the whole diseased tissue. However, in practice, each defective network only simulates one sort of defective cell or one subpopulation. We need an ensemble of networks to simulate the total cancer population, where the number of required networks equals the number of significant subpopulations in the cancer tissue. This ensemble of networks has to be determined from the data using professional expertise. In our new parallelizable model, the ensemble's subpopulations or networks influence the observables in a weighted average fashion. Our goal is to determine how much the behavior of the tissue is affected by watching the output behavior, which is dictated by the model parameters.
\subsection{An example}
\begin{figure}[htbp]
    \includegraphics[width=10cm, height=15cm]{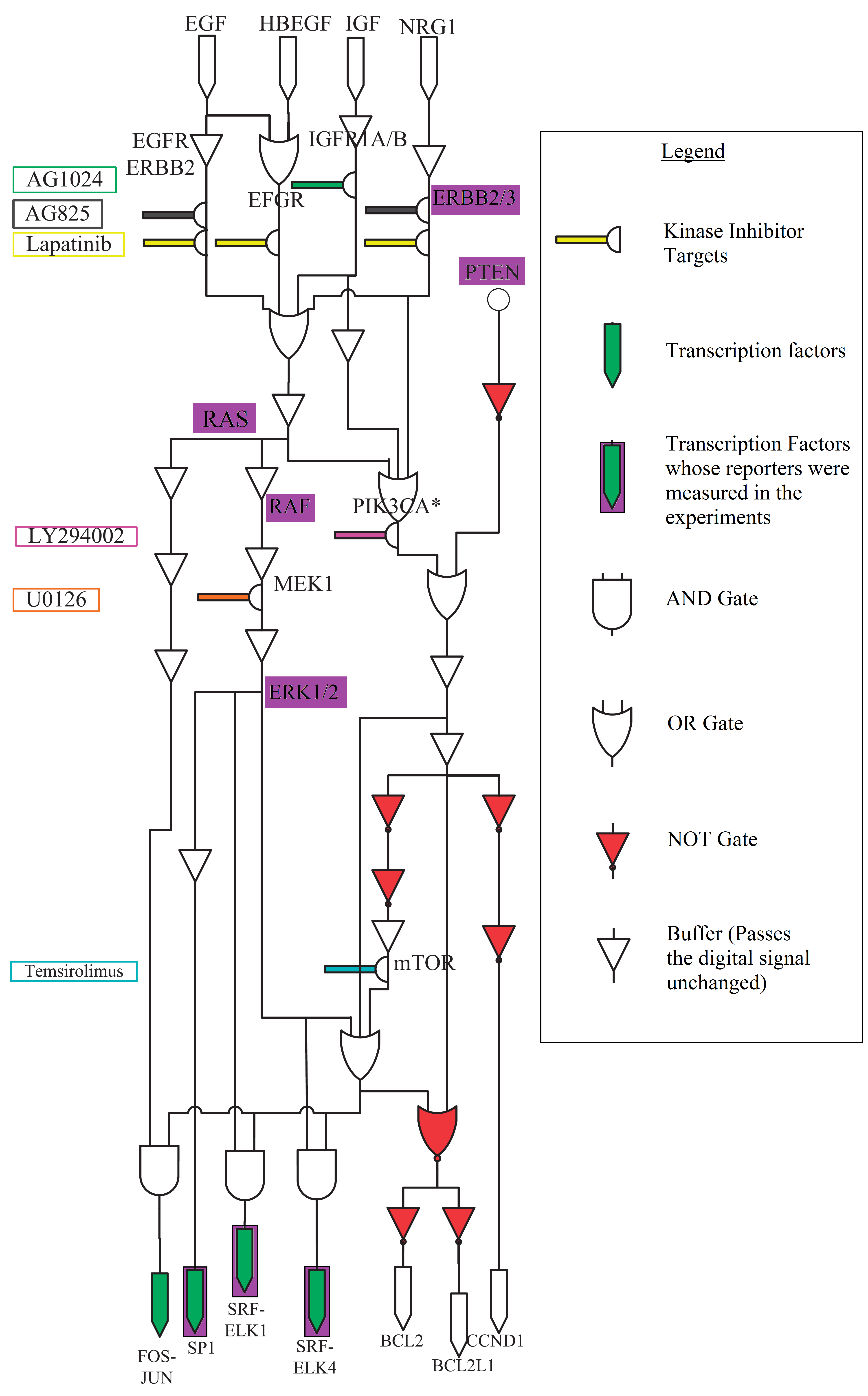}
    \caption{Boolean network model of the MAPK transduction network
             with target locations of drugs as described in \cite{18}. }
    \label{fig1}
\end{figure}

The outputs of the Boolean network shown in figure \ref{fig1} change depending on the exposed stimuli (a combination of kinase inhibitory drugs). The outputs can be one or zero, depending on the stimulus. One represents up$-$regulated, and zero represents down$-$regulated. The outputs can be different for each network in the collection. For example, if we consider three subpopulations, then there are three networks. Let us consider that the first subpopulation has a stuck$-$at one fault at ERK1/2, the second population has two stuck-at-one faults at ERBB2/3 and Raf, and the third population has a stuck-at-zero fault at PTEN. The purple rectangles in figure \ref{fig1} show the fault locations. Now, let us consider that the cell culture was exposed to U1026. U1026 is a kinase inhibitor drug and targets MEK1, as shown in figure \ref{fig1}. Targets of all the drugs are shown in figure \ref{fig1}, and all the drugs are kinase inhibitors. Assuming the serum contains EGF, HBEGF, IGF, and NRG1, and if SP1 is observed, the first network has upregulated SP1 or 1, and the second and third networks have downregulated SP1 or 0. By observing the expression of gene cMYC, we can observe the activity of SP1. The weight of the network would determine the extent to which the gene cMYC is affected. \par

The activities of the Boolean networks with respect to the $i^{th}$ gene can be represented by a vector $d_i = (1\; 0\; 0)^T$ in this example (upregulated SP1 for the first network and downregulated SP1 for the second and third network). Vector $d_i$ is defined as expression profile \cite{g}. The expression profile depends on the drugs given, faults in the network, the stimulus it is subjected to, and the number of networks present in the collection. The $d_i$s come with the data.\par

\subsection{GPU basics}
Graphics processing units follow the SIMD (Single Instruction Multiple Data) model of Computation, whereas CPUs follow the SISD (Single Instruction Single Data) model of computation; GPUs are capable of handling millions of computations in parallel, unlike a CPU built to run sequential calculations. GPUs can achieve this feat because it has thousands of cores optimized to handle multiple calculations in parallel. Computations on a GPU are done on threads; each thread in a thread block runs the same instruction simultaneously. Multiple thread blocks can be executed concurrently. GPU-compatible models are capable of reaping the benefits of GPU's parallelism. Therefore, if a problem can be mapped to the SIMD framework of GPU, it is clear that the speed of GPU computation would outperform the speed of CPU computation..\par
\emph{CUDA, OpenCL, HIP} are the three major GPU frameworks. CUDA is a popular framework for Nvidia GPUs and has a huge user base. CUDA libraries include \emph{cuBLAS} for linear algebra, \emph{cuRAND} for random number generation, and \emph{THRUST} to run Standard Template Library (STL) algorithms on a GPU. \par
The function running on a GPU is called a \emph{kernel}. Kernels are called from a host to run a code on the GPU. Kernel launch requires the number of thread blocks and threads per block to be specified. Threads can be identified by a unique identification number called \emph{threadIdx}.\par
We shall use \emph{cuRAND} to generate random numbers, \emph{cuBLAS} to find the inverse of matrices and to perform general matrix multiplication in batches, and \emph{THRUST} for reduction and other STL algorithms on GPU. Our code is written in CUDA C++.
It is worth noting that our subpopulation computation problem can be mapped to the SIMD architecture of a GPU.

\section{Methods}
In this paper, we consider cancer as stuck-at faults, and the faulty networks are chosen based on prior knowledge. After the networks are chosen, the problem is to calculate the weights related to each network from the data collected. Our methods use normalized gene expression ratios to estimate the weights of the subpopulation. A method to measure normalized gene expression reading from qPCR data is shown in \cite{8}.\par

In \cite{h}, authors used a Boolean network to analyze heterogeneity in cancer tissue. An ensemble of Boolean networks was considered as subpopulations, and its effect on the observables was represented as a weighted average. The authors in \cite{h} presented a model suitable for the parallel computing environment to calculate the weights linked with the subpopulation in the collection of networks. Metropolis$-$Hastings, which is a Markov Chain Monte Carlo (MCMC) algorithm, was used to solve this problem.  The method used in \cite{h} is computationally intensive, and it is difficult to judge convergence. Moreover, the method requires further thinning to obtain a good effective sample size. So, to overcome these problems, in this paper, we describe a parallelizable model that allows us to use variational Bayes to calculate the posterior distribution of the unknown weights on a GPU. Authors in \cite{20}, have used variational methods to the conjugate exponential family model. Therefore, we approximate the parallelizable cancer tissue heterogeneity model in \cite{h} by a conjugate exponential model and derive the update equations that are compatible with the Single Instruction Multiple Data (SIMD) architecture of GPUs.\par

\subsection{Short summary of previous work}
A parallelizable model for cancer tissue heterogeneity was first discussed in \cite{h}. Our work is a continuation of \cite{h}. In \cite{h}, the authors describe a parallelizable model to compute the subpopulation weights on a GPU. Normalized gene expression ratio can be modeled by a ratio of two normal distributions, each with its own mean and the standard deviation being proportional to the mean \cite{g} and \cite{28}. The normalized gene expression $r_i$ for the $i^{th}$ gene was modeled as a ratio of two normal distributions in \cite{h}. In \cite{h}, the conditional probability distribution of $r_i$ was derived as:

\begin{equation}\label{eq1}
\begin{split}
    P(r_{i}/\alpha_{i},d_{i},c) & = \frac{m_{i}(r_{i}+m_{i})}{\sqrt{2\pi}c(r_{i}^2+m_{i}^2)^\frac{3}{2}}
    \times exp\Bigg(-\frac{1}{2c^2}\frac{(r_{i}-m_{i})^2}{r_{i}^2+m_{i}^2}\Bigg)
\end{split}
\end{equation}

where $m_i = d_i^T\alpha_i$, $d_i$ is the expression profile, $\alpha_i$ represent the subpopulation weights, $r_i$ represents the gene expression measurements and $c$ is the coefficient of variation. The activities of the Boolean networks with respect to the $i^{th}$ gene were represented by a vector $d_i$ in \cite{h}. An explanation of $d_i$ has been given in section 2. The gene expression data $r_i$ and the expression profile $d_i$ are the input data. Equation \ref{eq1} shows that gene expression measurement depends on subpopulation weights $\alpha_i$ and the coefficient of variation $c$.  The Bayesian network representing the dependencies in \cite{h} is shown in figure 2.\par
In Figure 2, $V$ number of gene expression measurements were considered and every $\alpha_{i}$ is connected to one gene expression measurement $r_i$. All the components of $\alpha_i$ was constrained to be non negative and all the components of each $\alpha_i$ sums up to one. $K$ is the parameter vector of the Dirichlet distributed $\alpha_i$s. The unknown parameter $K$ and $c$ were learnt from the observed data $r_i$. A detailed description of the probability model is discussed in \cite{h}.\par

\subsection{Comparison with existing work}
\label{comp}
The models in \cite{g} and \cite{8} are not compatible with the parallel computing environment of a GPU because each $\alpha_i$ is associated with $n_i$ number of gene expression measurements as shown in figure 2. The time complexity of these models is $O(n_V)$, where $n_V$ represents the total number of gene expression measurements. The computation time of these models increases with the number of input data. \par

\noindent The authors in \cite{h} used a MCMC algorithm called Metropolis-Hastings algorithm on a GPU to learn the unknown parameters. The Bayesian network \cite{25} of the parallelizable model described in \cite{h} is shown in figure 2. However, M$-$H algorithm is computationally intensive, difficult to judge convergence and requires thinning to achieve good effective sample size, which further increases computation time.\par

\noindent In contrast to \cite{g} and \cite{8}, our model considers only one gene expression measurement for each $\alpha_i$ as shown in figure 3. Therefore our model can benefit from the parallelism offered by the SIMD environment of a GPU. The time complexity of our model is $i$, where $i$ is the number of Variational Bayes iterations; this is possible because all the computations in our model are suitable for the parallel computing environment of a GPU as shown in figure 3. Table 2 compares the time complexity of all the models.\par

To overcome the problems in \cite{h}, in this paper we show a new parallelizable model which allows us to run variational Bayes to compute the proportional breakup of heterogeneous cancer tissue on a GPU. In \cite{h}, the authors modeled the unknown parameters by unknown distributions, which lead to the usage of a computationally intensive algorithm called Metropolis-Hastings to sample from the posterior distributions of the unknown parameters. In this paper, we approximate the unknown parameters by a known distribution so that the posterior of the unknown parameters come out as a known distribution. Let $\alpha_i=(\alpha_{i,1}\; \alpha_{i,2}\; \alpha_{i,3})^T$ be the weights associated with each of the three Boolean networks discussed in section 2.  Authors in \cite{g}, \cite{h}, and \cite{29} have shown that the conditional distribution of gene expression measurement depends on $\alpha_i$ and a coefficient of variation. Therefore, to model the dependency of normalized gene expression measurements $r_i$ on $\alpha_i$ and the coefficient of variation $\rho$ we shall consider a multilevel hierarchical model, because a hierarchical model allows us to model such dependency. The normalized gene expression ratio $r_i$ depends on $\alpha_i$ and $\rho$, as shown in figure 3. The variable $\alpha_i$ depends on $K$ and $\Lambda$ shown in figure 3. The unknown parameters which need to be estimated are $\Lambda$, $K$, and $\rho$ shown in figure 3. The parameter $K$ is of particular interest because it represents the averaged view of the weights exerted by each Boolean network on the observables. The Bayesian network of our new parallelizable model is shown in figure 3. The unknown parameters $K$, $\rho$ and $\Lambda$ are learnt from the observed gene expression data $r_i$, here $r_i$ represent the gene expression measurement for the $i^{th}$ gene. 

\begin{center}
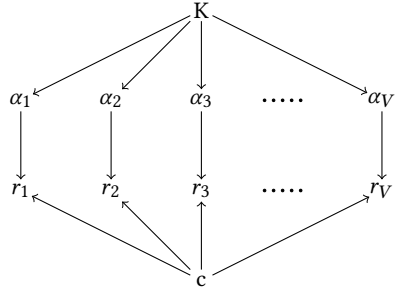

\begin{tikzpicture}[scale=1.2, auto,swap]
\filldraw [black] (.7,3) circle (.4pt);
\filldraw [black] (.8,3) circle (.4pt);
\filldraw [black] (.9,3) circle (.4pt);
\filldraw [black] (1,3) circle (.4pt);
\filldraw [black] (1.1,3) circle (.4pt);

\filldraw [black] (.7,2) circle (.4pt);
\filldraw [black] (.8,2) circle (.4pt);
\filldraw [black] (.9,2) circle (.4pt);
\filldraw [black] (1,2) circle (.4pt);
\filldraw [black] (1.1,2) circle (.4pt);
\foreach \pos/\name/\disp in {
  {(0,4)/1/K}, 
  {(-2,3)/2/$\alpha_1$}, 
  {(-1,3)/3/$\alpha_2$}, 
  {(0,3)/4/$\alpha_3$}, 
  {(2,3)/5/$\alpha_V$},
  {(2,2)/6/$r_V$},
  {(-1,2)/7/$r_2$},
  {(-2,2)/8/$r_1$},
  {(0,2)/9/$r_3$},
  {(0,1)/10/c}}
\node[minimum size=8pt,inner sep=0pt] (\name) at \pos {\disp};
\foreach \source/\dest in {
  1/2,
  1/3,
  1/4,
  1/5,
  2/8,
  3/7,5/6,
  4/9,10/7,10/8,10/9,10/6}
\path[draw,thin,->] (\source) -- node {} (\dest);
\end{tikzpicture}
\captionof{figure}{\small{The Bayesian network \cite{25} representing dependencies of the model in \cite{h}. Here $r_i$ is data, $\alpha_i$ represents subpopulation breakup, $K$ and $c$ are learnt}}
\end{center}

\begin{center}
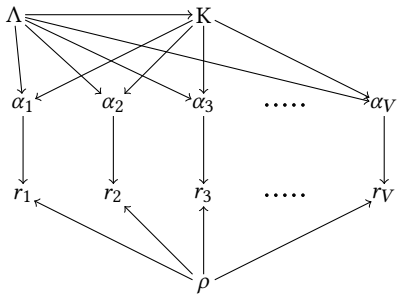

\begin{tikzpicture}[scale=1.2, auto,swap]
\filldraw [black] (.7,3) circle (.4pt);
\filldraw [black] (.8,3) circle (.4pt);
\filldraw [black] (.9,3) circle (.4pt);
\filldraw [black] (1,3) circle (.4pt);
\filldraw [black] (1.1,3) circle (.4pt);

\filldraw [black] (.7,2) circle (.4pt);
\filldraw [black] (.8,2) circle (.4pt);
\filldraw [black] (.9,2) circle (.4pt);
\filldraw [black] (1,2) circle (.4pt);
\filldraw [black] (1.1,2) circle (.4pt);
\foreach \pos/\name/\disp in {
  {(-2.1, 4)/0/$\Lambda$},
  {(0,4)/1/K}, 
  {(-2,3)/2/$\alpha_1$}, 
  {(-1,3)/3/$\alpha_2$}, 
  {(0,3)/4/$\alpha_3$}, 
  {(2,3)/5/$\alpha_V$},
  {(2,2)/6/$r_V$},
  {(-1,2)/7/$r_2$},
  {(-2,2)/8/$r_1$},
  {(0,2)/9/$r_3$},
  {(0,1)/10/$\rho$}}
\node[minimum size=8pt,inner sep=0pt] (\name) at \pos {\disp};
\foreach \source/\dest in {
  0/1,
  0/2,
  0/3,
  0/4,
  0/5,
  1/2,
  1/3,
  1/4,
  1/5,
  2/8,
  3/7,5/6,
  4/9,10/7,10/8,10/9,10/6}
\path[draw,thin,->] (\source) -- node {} (\dest);
\end{tikzpicture}
\captionof{figure}{\small{A Bayesian network representing the conditional dependencies of our parallelizable model (subsection 3.3)}}
\end{center}

\subsection{A Parallelizable Conjugate Exponential Model for Cancer Tissue Heterogeneity}
\begin{figure}[htbp]
\centering
    
\begin{tikzpicture}[node distance=2cm]

\node (start) [process] {\textbf{Data modeling:} Modeling of gene expression measurement 
and the weights by normal distribution (equations \ref{eq3} and \ref{eq4})};
\node (pro1) [process, below of=start] {\textbf{Assigning priors:} Consider a prior distribution over the unknown parameters (equations \ref{eq5}, \ref{eq6} and \ref{eq7})};


\node (stop) [process, below of=pro1] {\textbf{Factorizing the posterior:} Factorize the joint posterior distribution of the unobserved variables (equation \ref{eq8})};

\node (stop1) [process, below of=stop] {\textbf{Accelerating variational Bayes:} Use variational Bayes to compute the update equations on a GPU (subsection \ref{update})};

\draw [arrow] (start) -- (pro1);
\draw [arrow] (pro1) -- (stop);
\draw [arrow] (stop) -- (stop1);
\end{tikzpicture}

\caption{Block diagram explaining the computation of the unknown parameters for our proposed model by using variational Bayes} \label{block}
\end{figure}
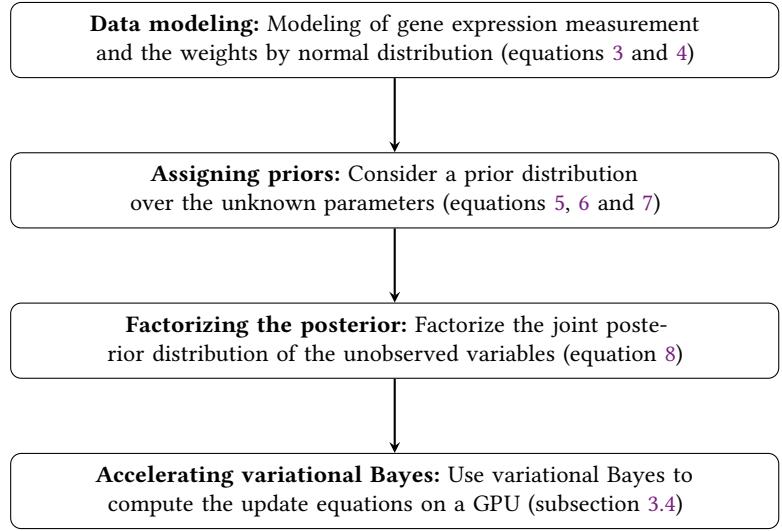

Figure 3 shows the Bayesian network of our new parallelizable model. Here, the weight vectors $\alpha_i$ run from 1 through $V$. Each $\alpha_i$ has $N$ components, where $N$ is the number of subpopulations. The number of boolean networks is equal to the number of subpopulations. For each gene, we make a single observation. Each observation is represented by $r_i$. There are $V$ number of genes, so there are $V$ number of gene expression measurements. Therefore, $r_i$ runs from 1 through $V$. Each $r_i$ is connected to one $\alpha_i$.  Every $r_i$ is considered to be normally distributed with mean $d_i^T\alpha_i$ and precision of $\rho$, which is the variance of $\rho^{-1}$. Here $d_i$ is the expression profile related to each $r_i$. Expression profile $d_i$ depends on the tissue stimulus and the networks in the collection. The expression profile comes with the observed data. The $\alpha_i$s come from a multivariate normal distribution with mean $K$ and a precision matrix of $\Lambda$, which is a covariance matrix of $\Lambda^{-1}$. The unknown parameters which need to be determined are $\Lambda$, $K$, and $\rho$. $K$ represents the weights of each Boolean network in the collection, so $K$ is our parameter of interest. $\alpha_i$s represent the weights exerted on the gene expression values for each gene by each Boolean network and vary from gene to gene. The averaged view of the weights exerted by each network in the collection is represented by the parameter $K$. Figure \ref{block} shows a block diagram of the entire process.\par

Referring to the data modeling block in figure \ref{block}, we have adapted the approach of \cite{g}, \cite{h}, and \cite{29} to model the gene expression measurements and the weights associated with each subpopulation by a normal distribution for our proposed model. The prior distributions over the unknown parameters are assigned as per \cite{21}, \cite{22}, and \cite{26} so that the posterior distribution of the unknown parameters result in known distributions. The priors over $K$, $\Lambda$ and $\rho$ are chosen to be normal, Wishart and Gamma distribution respectively. Variational Bayes \cite{20} assumes that the joint distribution can be factored, as shown in equation \ref{eq8}. Next we accelerate the variational Bayes algorithm on a GPU to compute the weightage of each subpopulation, i.e. the weight of each Boolean network on the observables. The acceleration of variational Bayes and the derivation of the posteriors of unknown parameters is discussed in subsection 3.4.\\
Given that we represent the effect of each Boolean network on the observables as a ratio, we reduce the redundancy by constraining the weight vectors $\alpha_i$ by

\begin{equation}\label{eq2}
\displaystyle\sum_{q=1}^N \alpha_{i,q} =1
\end{equation}

where $N$ is the number of subpopulations, which is equal to the number of Boolean networks in the collection and each $\alpha_i$ has $N$ components, so $\alpha_{i,q}$ represent the $q^{th}$ value of $\alpha_i$. Therefore, $\alpha_{i,N}$ can be represented by $\alpha_{i,N} = 1 - \sum_{q=1}^{N-1}\alpha_{i,q}$. Following the discussion in \cite{g} and \cite{h} we can assume the gene expression reading $r_i$ to be normally distributed with mean $d_i^T\alpha_i$ and $\rho$ as the precision. Following the discussion in \cite{g} and \cite{h}, the mean can be defined from equation \ref{eq2} as $d_i^T\alpha_i = \sum_{q=1}^{N-1}\alpha_{i,q}(d_{i,q}-d_{i,N})+d_{i,N}$. Let us define $d_{i,q}$ as the elements of the expression profile $d_i$.\\
Defining $\mu_i$, $D_i$ and $\beta_i$ to be $\mu_i = d_{i,N}$, $D_i = (d_{i,1}-d_{i,N},d_{i,2}-d_{i,N},\dots,d_{i,N-1}-d_{i,N})^T$, and $\beta_i =(\alpha_{i,1},\alpha_{i,2},\dots,\alpha_{i,N-1})^T$, the probability distribution of gene expression $r_i$ for the $i^{th}$ gene is

\begin{equation}\label{eq3}
P(r_i|\beta_i,\rho,d_i) = \mathcal{N}(r_i|D_i^T\beta_i+\mu_i,\rho^{-1})
\end{equation}

The probability distribution of $\beta_i$ is

\begin{equation}\label{eq4}
P(\beta_i|K,\Lambda) = \mathcal{N}(\beta_i|K,\Lambda^{-1})
\end{equation}
where $K=(K_1\; K_2\;\dots\; K_{N-1})^T$ is the mean and $\Lambda$ is a precision matrix. $\Lambda$ is a positive definite matrix, it is an $(N-1)\times(N-1)$ matrix;
the unknown parameters $\rho$, $K$ and $\Lambda$ are called unobserved variables in the Bayesian framework. The weights of $N-1$ networks can be represented by $K$. Weight of the $N^{th}$ network can be found out by the expression $1-\sum_{q=1}^{N-1}K_q$.\par
According to the Bayesian point of view, we have to define few priors over the unknown variables. So, a Gamma distribution with shape and inverse scale parameters $a_0$ and $b_0$ is considered as a prior over $\rho$. We consider the prior over $K$ to be a normal distribution, and the prior over $\Lambda$ as a Wishart distribution. These priors are chosen so that the posterior comes out to be a known distribution \cite{20} and \cite{21}. Therefore, the priors are

\begin{equation}\label{eq5}
P(\rho) = Gamma(\rho|a_0,b_0)
\end{equation}

\begin{equation}\label{eq6}
P(K/\Lambda) = \mathcal{N}(K|K_0,(q_0\Lambda)^{-1})
\end{equation}

\begin{equation}\label{eq7}
P(\Lambda) = Wish(\Lambda|n_0,(\Lambda_0)^{-1})
\end{equation}
The distributions $P(r_i|...)$, $P(\beta_i|...)$ shown in equation \ref{eq3} and \ref{eq4} and the prior distribution over the unknown parameters shown in equation \ref{eq5}, \ref{eq6} and \ref{eq7} are chosen such that the distribution of the unknown parameter comes out to be a known distribution after using the variational Bayes, expectation maximization and the Markov chain Monte Carlo algorithm \cite{20}, \cite{21} and \cite{25}.
\noindent Here $K_0,\;q_0,\;n_0$ and $\Lambda_0$ are constants.\\
The joint posterior distribution of the unknown variables $\rho, \Lambda, K$ is

\begin{equation}\label{eq8}
\begin{split}
   P(\rho,\beta,K,\Lambda|r)& \propto P(\Lambda)P(K/\Lambda)P(\rho)\times
   \prod_{i=1}^{V}\Big(P(r_i/\beta_i,\rho,d_i)P(\beta_i/K,\Lambda)\Big)
\end{split}
\end{equation}

Here $\beta$ is a set of all the $\beta_i$s shown above and $r$ is the set of all the observed gene expression measurements $r_i$. The hierarchical model that we considered here belongs to the family of conjugate exponential \cite{21} and \cite{25}. In a conjugate exponential model, the conditional probability distribution belongs to the exponential distribution family, and they belong to the same family as their prior distribution. We approximate the joint distribution of the parameters of interest (for example $K$ which represents the proportional breakup of the heterogeneous cancer tissue) by utilizing variational method. This would help us to derive the marginal distributions of the unknown variables by using variational Bayes \cite{21}. In the next section, we derive the variational update equations.\par 
The joint posterior distribution of the unobserved variables can be assumed by a factorized form by using variational Bayes method. For example, we want to approximate $P(Z/D)$ by $Q(Z)$, where $Z$ is the set of unobserved variables. $P(Z/D)$ is the posterior distribution of the unobserved variables conditional on the data $D$. Then $Q(Z)$ can be represented by

\begin{equation}\label{eq9}
Q(Z) = \prod_{i=1}^M Q_i(Z_i)
\end{equation}
$Z_1$ through $Z_M$ are the disjoint partitions of the unobserved variables $Z$ \cite{21}. Next, the Kullback$-$Leibler (KL) divergence $KL(Q(Z)||P(Z/D)$ is minimized. It is important to note that

\begin{equation}\label{eq10}
\ln P(D) = KL(Q(Z)||P(Z/D)) + L_Q(D)
\end{equation}
where $L_Q(D)$ can be represented by

\begin{equation}\label{eq11}
L_Q(D) = \int \ln\Big(\frac{P(Z,D)}{Q(Z)}\Big)Q(Z)dZ
\end{equation}

Readers can refer to \cite{21} for a detailed derivation of equation \ref{eq10}. $L_Q(D)$ is called the lower bound. The following update equation minimises KL divergence.

\begin{equation}\label{eq12}
\ln Q_j(Z_j)=\int \ln P(Z,D)\prod_{i\neq j }Q_i(Z_i)dZ_i + constant
\end{equation}
A detailed derivation of equation \ref{eq12} is presented in \cite{21}. In the next section, we shall solve the equations to find the optimal approximations of the unobserved variables compatible with the parallel computing environment. Our parallelizable model allows us to compute the proportional breakup of heterogeneous cancer tissue on a GPU by using variational Bayes method.

\subsection{Derivation of the variational update equation}
\label{update}
We shall now develop the update equations of the unknown parameters by using variational Bayes algorithm. Here, $K$, $\rho$ and $\Lambda$ are the unknown parameters. The parameter $K$ is of interest, because it represents the averaged weight exerted by each Boolean network on the data. Let $Q(\rho, \beta, K,\Lambda)$ be the approximation for the posterior distribution $P(\rho,\beta,K,\Lambda/r)$. Assuming the approximation factors into

\begin{equation}\label{eq13}
  Q(\rho, \beta, K,\Lambda) = Q_\rho(\rho)Q_\beta(\beta)Q_{K,\Lambda}(K,\Lambda)
\end{equation}
 Using equation \ref{eq12} to find $Q_\rho(\rho)$ and by absorbing any term not dependent on $\rho$ to the $constants$ we have.

\begin{equation}\label{eq14}
\begin{split}
    \ln Q_\rho(\rho)=E_{\neq\rho}\Big[\ln P(\rho)+\sum_{i=1}^V\ln P(r_i/\beta_i,\rho,d_i)\Big]+constants
\end{split}
\end{equation}

Simplifying equation \ref{eq14} we get

\begin{equation}\label{eq15}
 \ln Q_\rho(\rho)=a_\rho\ln(\rho) -b_\rho\rho+constants
\end{equation}
where $a_\rho$ is represented by

\begin{equation}\label{eq16}
\begin{split}
 a_\rho = a_0 + \frac{1}{2}n_V
\end{split}
\end{equation}
here, $n_V=V$ is the total number of gene expression measurements and $b_\rho$ is represented by

\begin{equation}\label{eq17}
\begin{split}
     b_\rho =  b_0+
     \frac{1}{2}\sum_{i=1}^{V}\{(r_i-\mu_i)^2-&2(r_i-\mu_i)D_i^TE[\beta_i]
     +D_i^T(E[\beta_i\beta_i^T]D_i)\}
\end{split}
\end{equation}
By observing equation \ref{eq15}, \ref{eq16} and \ref{eq17} we deduce that $Q_\rho(\rho)$ is Gamma distributed with shape parameter $a_\rho$ and inverse scale parameter $b_\rho$. Therefore,

\begin{equation}\label{eq18}
Q_\rho(\rho)=Gamma(\rho|a_\rho,b_\rho)
\end{equation}

Similarly applying equation \ref{eq12} to derive the update equations for $Q_\beta(\beta)$ we get $Q_\beta(\beta) = \prod_{i=1}^{V}Q_{\beta_i}(\beta_i)$. After simplifying, $Q_{\beta_i}$ appears to be normally distributed with parameters $\mu_{\beta_i}$ and $\Lambda_{\beta_i}^{-1}$ as shown in \ref{eq19}, \ref{eq20} and \ref{eq21}.

\begin{equation}\label{eq19}
Q_{\beta_i}(\beta_i)=\mathcal{N}(\beta_i|\mu_{\beta_i},\Lambda_{\beta_i}^{-1})
\end{equation}
where $\Lambda_{\beta_i}$ can be represented as

\begin{equation}\label{eq20}
\Lambda_{\beta_i} = E[\Lambda] + E[\rho]D_{i}D_i^T
\end{equation}

and $\mu_{\beta_i}$ can be represented as

\begin{equation}\label{eq21}
\mu_{\beta_i}=\Lambda_{\beta_i}^{-1}\{E[\Lambda K]+E[\rho]D_i (r_i-\mu_i)\}
\end{equation}

$\Lambda_{\beta_i}$ is a positive semidefinite matrix with $(N-1)$ rows and $(N-1)$ columns. $\mu_{\beta_i}$ is a mean vector with length $(N-1)$.\\
Following a similar approach, the distribution of $Q_{K,\Lambda}(K,\Lambda)$ and $Q_{K,\Lambda}(\Lambda)$ comes out to be a normal and a Wishart distribution respectively as shown in equation \ref{eq22} and \ref{eq23}. The parameters of these distributions are shown in equations \ref{eq24} and \ref{eq25}.

\begin{equation}\label{eq22}
Q_{K,\Lambda}(K,\Lambda)=\mathcal{N}(K|K_{0K},[(q_0+V)\Lambda]^{-1})
\end{equation}

\begin{equation}\label{eq23}
Q_{K,\Lambda}(\Lambda)=Wish(\Lambda|n_0+V,\Lambda_{0\Lambda}^{-1})
\end{equation}

where $K_{0K}$ is defined as

\begin{equation}\label{eq24}
K_{0K} = \frac{\sum_{i=1}^V E[\beta_i]+q_0K_0}{V+q_0}
\end{equation}

and $\Lambda_{0\Lambda}^{-1}$ is represented as

\begin{equation}\label{eq25}
\begin{split}
    \Lambda_{0\Lambda}^{-1}=\Lambda_0^{-1}+\sum_{i=1}^V&E[\beta_i\beta_i^T]
    +q_0K_0K_0^T-(q_0+V)K_{0K}K_{0K}^T
\end{split}
\end{equation}
where $K_{0K}$ is a vector of length $(N-1)$ and $\Lambda_{0\Lambda}$ is a matrix of dimension $(N-1)\times(N-1)$.\par
We have adapted the approach discussed in \cite{20} and \cite{21} to derive the expectations for our proposed model, the expectations are:

\begin{equation}\label{eq26}
E[\beta_i]=\mu_{\beta_i}
\end{equation}

\begin{equation}\label{eq27}
E[\beta_i\beta_i^T]=\mu_{\beta_i}\mu_{\beta_i}^T+\Lambda_{\beta_i}^{-1}
\end{equation}

\begin{equation}\label{eq28}
E[\Lambda]=(n_0+V)\Lambda_{0\Lambda}
\end{equation}

\begin{equation}\label{eq29}
E[\rho]=\frac{a_\rho}{b_\rho}
\end{equation}

\begin{equation}\label{eq30}
E[K]=K_{0K}
\end{equation}

\begin{equation}\label{eq31}
E[\Lambda K]=(n_0+V)\Lambda_{0\Lambda}K_{0K}
\end{equation}

As described in the previous section, the parameter $K$ is of interest because it represents the averaged view of the weights exerted by each network. Since $K$ is normally distributed with parameters $K_{0K}$ and $\Lambda$ as shown in equation \ref{eq22}, so these parameters need to be computed before sampling $K$. $\Lambda$ is Wishart distributed with parameters $\Lambda_{0\Lambda}^{-1}$, $n_0$ and $V$ as shown in equation \ref{eq23}, where $\Lambda_{0\Lambda}^{-1}$ depends on $E[\beta_i\beta_i^T]$ and $K_{0K}$ as shown in equation \ref{eq25}. Equation \ref{eq27} shows that $E[\beta_i\beta_i^T]$ depend on $\mu_{\beta_i}$ and $\Lambda_{\beta_i}$. Equations \ref{eq28}, \ref{eq29}, \ref{eq30} and \ref{eq31} are used to compute $b_{\rho}$, $\Lambda_{\beta_i}$ and and $\mu_{\beta_i}$ respectively.

\subsection{Computing the unknown parameters with variational Bayes on a GPU}
\begin{figure}
     \centering
     \begin{subfigure}{0.4\textwidth}
         \centering
         \includegraphics[width=6.5cm, height = 6cm]{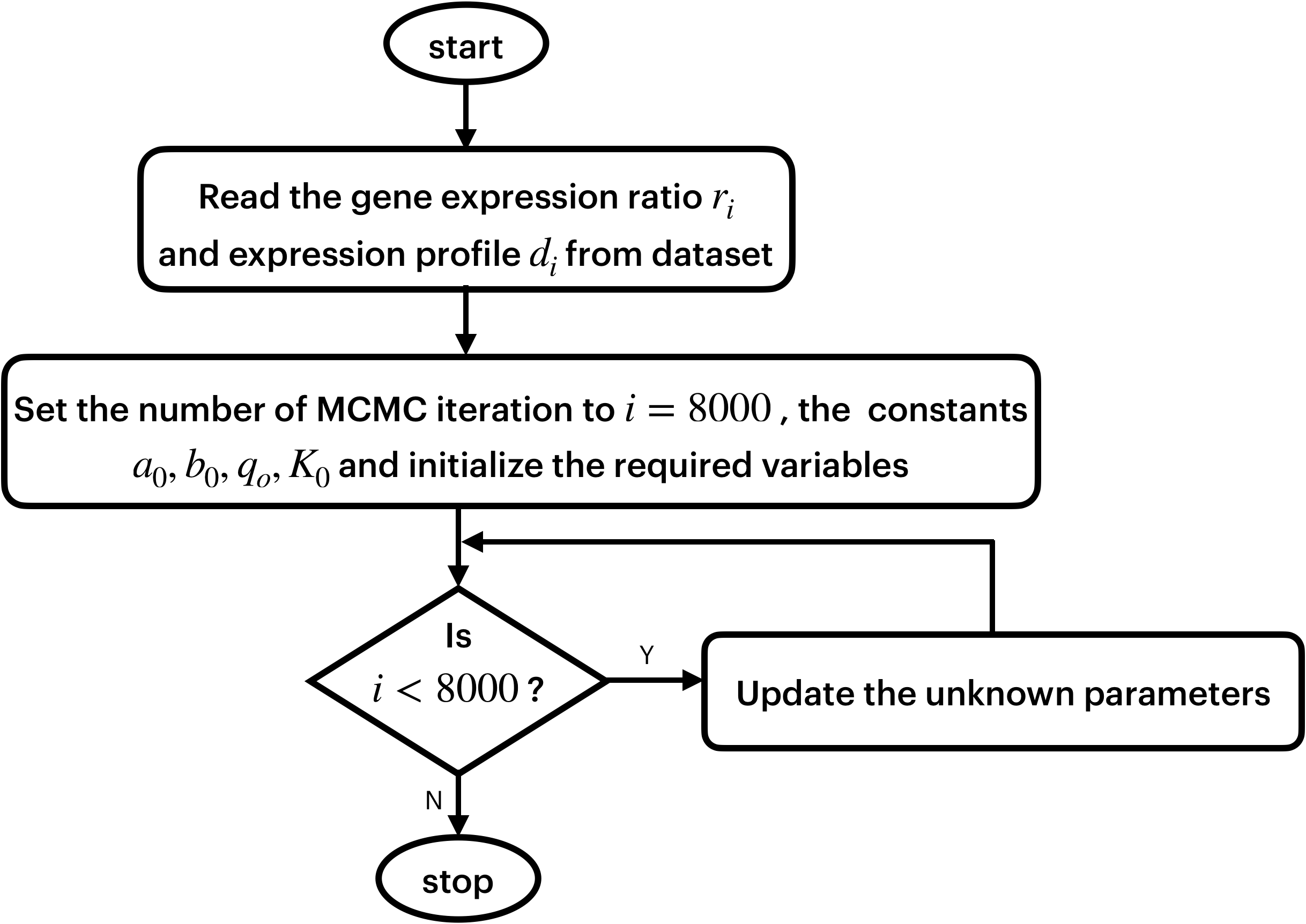}
         \caption{MCMC}
         \label{mcmc}
     \end{subfigure}
     \hfill
     \begin{subfigure}{0.4\textwidth}
         \centering
         \includegraphics[width=6cm, height = 6cm]{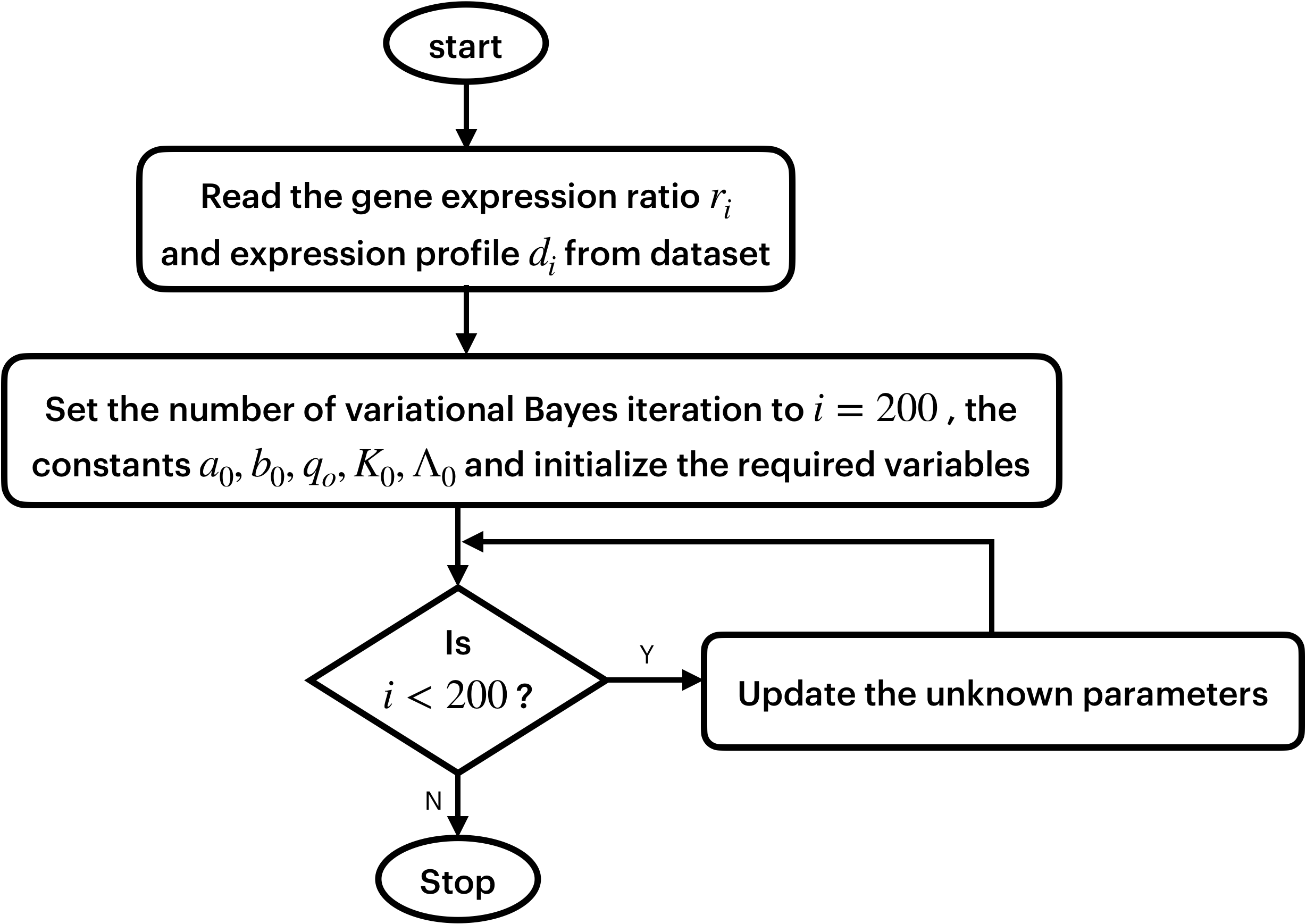}
         \caption{VB}
         \label{vb}
     \end{subfigure}
     \hfill
     \begin{subfigure}[b]{0.4\textwidth}
         \centering
         \includegraphics[width=6.5cm, height = 6cm]{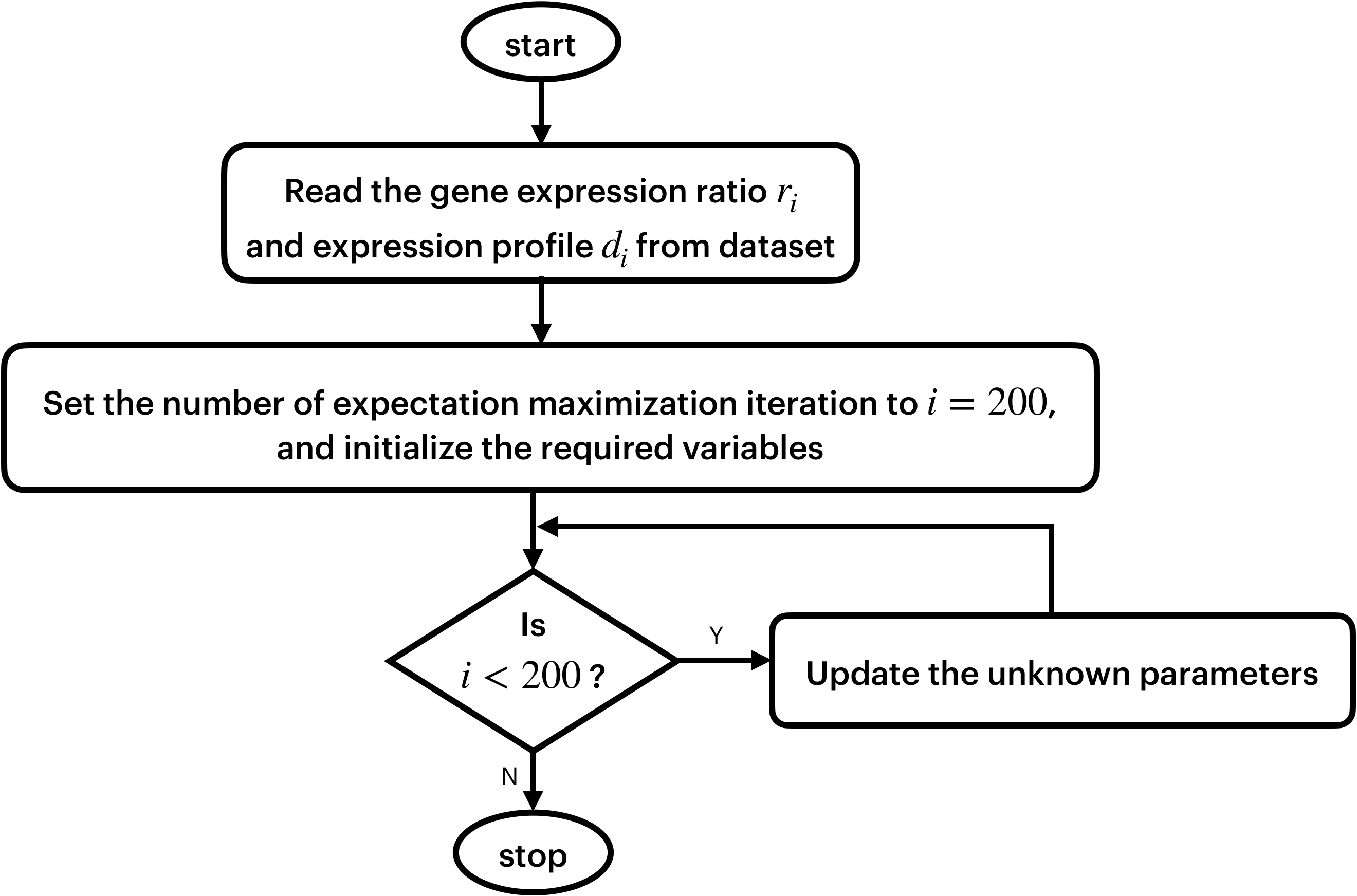}
         \caption{EM}
         \label{em}
     \end{subfigure}
        \caption{Flowchart showing the computation of the unknown parameters by using (a) MCMC, (b) variational Bayes and (c) expectation maximization method}
        \label{fig:three graphs}
\end{figure}

Our method computes the unobserved variables on a GPU by cycling through the update equations \ref{eq16} through \ref{eq25} and using equations \ref{eq26} through \ref{eq31} to calculate the expectations. The constants are set to appropriate values. $K_0$ is a vector and its elements are set to $(1/3\;1/3)^T$, $a_o$ and $b_0$ are both set to 0.5 , $q_0$ is set to 0.001 and $n_0$ is held constant at 1. $\Lambda_0$ is set to
\[\Lambda_0=\begin{bmatrix}
0.01 & 0.005\\
0.005 & 0.008\\
\end{bmatrix}^{-1}\]
Figure \ref{vb} presents a simplified explanation of the computations involved in the variational Bayes method.
 Referring to the flowchart of the variational Bayes method shown in figure \ref{vb}, we discuss the \emph{update the unknown parameters} block shown in the flowchart. The unknown parameter $K$ is of interest because it represents the averaged view of the weights exerted by each network. The objective is to compute the correct value of $K$ given the gene expression data $r_i$ and expression profile $d_i$. In this computation \emph{cublasDgemmBatched} (a function in \emph{cuBLAS}) was used to do batched matrix multiplication and \emph{cublasDmatinvBatched} (a function in \emph{cuBLAS}) was used to find the inverse of matrices in batches on a GPU.
The procedure to compute the subpopulation weights by using variational Bayes is as follows:
\begin{enumerate}[Step 1.]
    \item Compute  $E[\beta_i\beta_i^T]$
    \item Compute the inverse of $\Lambda_{0\Lambda}$ and then calculate $E[\Lambda]$
    \item Calculate $E[\Lambda K]$
    \item Use $E[\beta_i\beta_i^T]$ and calculate $b_\rho$
    \item Use $E[\Lambda]$, $E[\Lambda K]$ and $b_\rho$ calculated above, and compute $\Lambda_{\beta_i}$ and $\mu_{\beta_i}$
    \item Use $\mu_{\beta_i}$ computed above, and then compute $K_{0K}$. Use $K_{0K}$, and $E[\beta_i\beta_i^T]$ to compute  $\Lambda_{0\Lambda}^{-1}$
    \item Using the computed variables above, the unknown parameters $K$, $\Lambda$ and $\rho$ are sampled
\end{enumerate}

Now, we shall demonstrate the steps to find out the unknown parameters on a GPU by using Variational Bayes method. We shall show that these computations are well suited for the SIMD architecture of a GPU. $\Lambda_{\beta_i}$s and $\Lambda_{0\Lambda}$ are initialized to $\Lambda_0$. $\mu_{\beta_i}$s and $K_{0K}$ are intitalized to $(1/3\;1/3)^T$. $a_\rho$ and $b_\rho$ are initialized to $a_0$ and $b_0$. The following steps are followed to update $\mu_{\beta_i}$, $\Lambda_{\beta_i}$, $\Lambda_{0\Lambda}$, $K_{0K}$,  $a_\rho$ and $b_\rho$ on a GPU.

$\Lambda_{\beta_i}$s are saved in a vector of length $(N-1)\times (N-1) \times n_V$, where $n_V$ is the number of observed gene expression measurements. $\mu_{\beta_i}$s are stored in a vector of length $(N-1) \times n_V$.
 
\begin{enumerate}
\item \textbf{Parallel computation of $\mathbf{E[\beta_i\beta_i^T]}$}  \\
    \begin{enumerate}[(i)]
        \item Use $cublasDmatinvBatched$ to calculate the inverse of all the $\Lambda_{\beta_i}$ in parallel and store the result in a vector called $dinAb$, this vector is of length $(N-1)\times (N-1) \times n_V$
        \item Calculate $\mu_{\beta_i}$$\mu_{\beta_i}^T$ by using \emph{cublasDgemmBatched} and store it in a vector called $rdex_a$ of length $(N-1)\times (N-1) \times n_V$
        \item Add $rdex_a$ and $dinAb$ in a CUDA kernel and save the result in a vector of length $(N-1)\times (N-1) \times n_V$. This vector contains $E[\beta_i\beta_i^T]$
\end{enumerate}

\item \textbf{Computing $\mathbf{E[\Lambda]}$}
    \begin{enumerate}[(i)]
        \item Calculate the inverse of $\Lambda_{0\Lambda}$ by using \emph{cublasDmatinvBatched}
        \item Calculate $E[\Lambda]$ as shown in equation \ref{eq28}
    \end{enumerate}
\item $\mathbf{E[\Lambda K]}$ was calculated by using \emph{cublasDgemm}
\item \textbf{Calculating $\mathbf{b_\rho}$}:
\begin{enumerate}[(i)]
    \item Calculate $D_i^T(E[\beta_i\beta_i^T])$  and $D_i^T(E[\beta_i\beta_i^T])D_i$ by using \emph{cublasDgemmBatched}. $E[\beta_i\beta_i^T]$ was calculated in step 1
    \item Calculate $D_i^T(E[\beta_i\beta_i^T])D_i$ by using \emph{cublasDgemmBatched}. Store the result in a vector of length $n_V$
    \item Calculate $D_i^TE[\beta_i]$ by using \emph{cublasDgemmBatched}. Calculate equation \ref{eq17} by using $thrust::reduce$
\end{enumerate}
\item \textbf{Calculating $\mathbf{K_{0K}}$ and $\mathbf{\Lambda_{0\Lambda}^{-1}}$}:
\begin{enumerate}[(i)]
    \item Calculate $\sum_{i=1}^VE[\beta_i]$ by $thrust::reduce$. $K_{0K}$ is calculated as shown in equation \ref{eq24}
    \item Use \emph{thrust::reduce} to calculate $\sum_{i=1}^VE[\beta_i\beta_i^T]$. Use \emph{cublasDgemm} to calculate $q_0K_0K_0^T$ and $(q_0+V)K_{0K}K_{0K}^T$. Then, $\Lambda_{0\Lambda}^{-1}$ is calculated as shown in equation \ref{eq25}
\end{enumerate}
\item \textbf{Calculating} $\mathbf{\Lambda_{\beta_i}}$ and $\mathbf{\mu_{\beta_i}}$:
\begin{enumerate}[(i)]
    \item Equation \ref{eq20} is calculated by \emph{cublasDgemmBatched}
    \item The inverse of $\Lambda_{\beta_i}$ is calculated by \emph{cublasDmatinvBatched}. Then, $\mu_{\beta_i}$ is calculated by \emph{cublasDgemmBatched}
\end{enumerate}
\item \textbf{Sample $\mathbf{\Lambda}$ on a GPU}.\\ The following steps demonstrate the sampling procedure from a Wishart distribution \cite{24} on a GPU
\begin{enumerate}[(i)]
    \item Calculate the Cholesky decomposition of $\Lambda_{0\Lambda}^{-1}$ by using \emph{cusolverDnDpotrfBatched} and store the result in \emph{V}
    
    \item Sample $(n_0+V)(N-1)$ normally distributed samples by using \emph{curand\_normal}, and store them in a vector called \emph{u}. $V$ is copied $(n_0+V)$ times in a vector called \emph{R}
    
    \item Calculate $S = Ru$ by using \emph{cublasDgemmBatched}
    \item Calculate $\Lambda=\sum_{i=1}^{n_0+V}S_iS_i^T$ by \emph{cublasDgemmBatched} followed by \emph{thrust::reduce}
\end{enumerate}
\item \textbf{Sample $\mathbf{K}$}.\\ The following steps demonstrate the sampling procedure from a normal distribution on a GPU
\begin{enumerate}[(i)]
    \item Calculate the Cholesky decomposition (decomposes a positive definite matrix to lower triangular matrix and its conjugate transpose) of $[(q_0+V)\Lambda]^{-1}$ by using \emph{cusolverDnDpotrfBatched}. Store the decomposition in a vector called $R1$
    \item Sample $(N-1)$ normally distributed samples by using \emph{curand\_normal}, and store them in a vector called \emph{u1}
    \item Calculate $K=R1u1+K_{0K}$ by using \emph{cublasDgemm}
\end{enumerate}
\item \textbf{Sample $\mathbf{\rho}$}.\\
A custom built CUDA kernel was written to sample from $Gamma(a_\rho, b_\rho)$ by following the method in \cite{h}. The kernel is described as follows:
\begin{enumerate}[(i)]
    \item Compute $w=a_\rho$ and $z=b_\rho$. $b_\rho$ is computed by following the steps mentioned above.
        \item Calculate \textbf{$d_{2}=w-\frac{1}{3}$}
        \item Compute $c_2=\frac{1}{\sqrt{9*d_{2}}}$
        \item Draw $u_{1}$, a uniformly distributed sample, by using \emph{curand\_uniform} (\emph{curand\_uniform} is a sampler in CUDA)
        \item Draw $u_{23}$, a sample from normal distribution with mean 0 and standard deviation 1, by using \emph{curand\_normal} (\emph{curand\_normal} is a sampler in CUDA)
        \item Compute $v_2=(1+c_2*u_{23})^3$
        \item Check if\\ $(j_2 = v_2>0)$ and $log(u_{1})<\frac{1}{2}u_{23}^2+d_{2}-d_{2}*v_2+d_{2}*log(v_2)$. The result in $j_2$ will either be a 0 or 1
        \item If $j_2$ is equal to 1, then return $\frac{d_{2}*v_2}{z}$ as the sample from $Gamma(a_\rho, b_\rho)$. Else go to step iv
\end{enumerate}

\end{enumerate}

\section{Experiments with synthetic data}
Only one real experimental dataset \cite{8} was available to us. So, to have a better model validation we use synthetic data in addition to the experimental dataset. We considered three Boolean networks to generate the synthetic data, so $N=3$. The faults that we considered are stuck at one fault at Ras for the first network, a stuck at zero fault at PTEN for the second network and stuck at one fault at RAF for the third network. The fault locations are shown in figure 1. The activity of the transcription factors in figure 1 would depend on the given combination of drugs. There are six different drugs, so 63 drug combinations are possible, leaving out occurrence of no drugs. To demonstrate, we considered $V$ genes, and each gene is associated with one gene expression measurement as shown in figure 3. We considered $K=(0.1\; 0.3)^T$, which means the first, second and the third networks have weights $0.1$, $0.3$ and $0.6$ respectively.. We let $\rho = 100$ and $\Lambda$ be
\[\Lambda=\begin{bmatrix}
0.01 & 0.005\\
0.005 & 0.008\\
\end{bmatrix}^{-1}\]

To compare our results with the variational Bayes we found out the posterior distributions of the unknown parameters by using Gibbs sampling (a MCMC algorithm) and found the point estimates by using expectation maximization. We shall also run these two algorithms on a GPU to compare our results.

\subsection{Computing the sub-populations by using Gibbs sampler}
Here, $K$ is the parameter of interest, because it represents the weights exerted by each network. The full conditional distributions of the unknown parameters $\Lambda$, $K$ and $\rho$ are shown in equations \ref{eq32}, \ref{eq35} and \ref{eq36}. Here, we have assumed the same data modeling as shown in equations \ref{eq3}, \ref{eq4} and prior assignments shown in equations \ref{eq5}, \ref{eq6}, \ref{eq7}. 
The objective is to estimate the correct value of $K$ given the gene expression data $r_i$ and expression profile $d_i$. Following Bayesian rule, the posterior distribution of the unknown parameters i.e. $K$, $\rho$ and $\Lambda$ are computed. Gibbs sampler is a Markov Chain Monte Carlo (MCMC) algorithm that allows sampling from posterior distributions. \cite{24} has an exhaustive explanation on this method, so here we will focus on the usage of this method. To use Gibbs sampler, we need the full conditionals.  \\
The full conditional of $\Lambda$ is:
\begin{equation}\label{eq32}
    P(\Lambda|\dots) = Wish(\Lambda|,n_\Lambda,\Lambda_\Lambda^{-1})
\end{equation}
where
\begin{equation}\label{eq33}
    n_\Lambda=n_0+V+1
\end{equation}
and
\begin{equation}\label{eq34}
\begin{split}
    \Lambda_\Lambda=\Lambda_0^{-1}+q_0(K-K_0)&(K-K_0)^T+
    \sum_{i=1}^V(\beta_i-K)(\beta_i-K)^T
\end{split}
\end{equation}
The full conditional of $K$ is:
\begin{equation}\label{eq35}
   P(K|\dots)=\mathcal{N}\Bigg(K|\frac{q_0K_0+\sum_{i=1}^V\beta_i}{q_0+V},((q_0+V)\Lambda)^{-1}\Bigg)
\end{equation}
The full conditional of $\rho$ is:
\begin{equation}\label{eq36}
   P(\rho|\dots)=Gamma(\rho|a_\rho,b_\rho)
\end{equation}
where
\begin{equation}\label{eq37}
    b_\rho =  b_0+
     \frac{1}{2}\sum_{i=1}^{V}\{(r_i-\mu_i-D_I^T\beta_i)^2\}
\end{equation}
and
\begin{equation}\label{eqa}
    a_\rho = a_0+\frac{1}{2}n_V
\end{equation}
where $n_V$ is equal to $V$.\\
The full conditional of the $\beta_i$s are:
\begin{equation}\label{eq38}
    P(\beta_i|\dots)=\mathcal{N}(\beta_i|\mu_{\beta_i},\Lambda_{\beta_i}^{-1})
\end{equation}
where
\begin{equation}\label{eq39}
    \Lambda_{\beta_i}=\Lambda+\rho D_i D_i^T
\end{equation}
\begin{equation}\label{eq40}
    \mu_{\beta_i}=\Lambda_{\beta_i}^{-1}\Bigg(\Lambda K+\rho(r_i-\mu_i)D_i\Bigg)
\end{equation}
Here, $K_0$ is a vector and its elements are set to $(1/3\;1/3)^T$, $a_o$ and $b_0$ both are set to 0.5, $q_0$ is set to 0.001 and $n_0$ is held constants at 1 and $V$ is the number of genes.
Figure \ref{mcmc} explains the usage of an Markov chain Monte Carlo (MCMC) algorithm called Gibbs sampler to compute the unknown parameters.
The procedure to compute the unknown parameters by using Gibbs sampler is as follows:
\begin{enumerate}[Step 1.]
    \item Sample $\Lambda$ from a Wishart distribution
    \item Use the sampled $\Lambda$ and sample $K$ from normal distribution
    \item Use the sampled $K$ and $\Lambda$ to sample all $\beta_i$s from a multivariate normal distribution
     \item Use $\beta_i$s and sample $\rho$ from Gamma distribution
\end{enumerate}

The following steps describe the sampling procedure on a GPU.
\begin{enumerate}
    \item \textbf{Sample $\mathbf{\Lambda}$}
    \begin{enumerate}[(i)]
        \item Calculate $\sum_{i=1}^V (\beta_i-K)(\beta_i-K)^T$ by using \emph{cublasDgemmBatched} and then followed by \emph{thrust::reduce}. \emph{thrust::reduce} is used to reduce a vector.
        \item Calculate $q_0(K-K_0)(K-K_0)^T$ by using \emph{cublasDgemm}
        \item Equation \ref{eq34} is calculated by adding all the vectors previously calculated on a custom built CUDA kernel
        \item A sample from $Wish(\Lambda|,n_\Lambda,\Lambda_\Lambda^{-1})$ is drawn by following the method discussed in the section 3
    \end{enumerate}
    
    \item \textbf{Sample $\mathbf{K}$}
    \begin{enumerate}[(i)]
        \item Calculate $(q_{0}+V) \Lambda)^{-1}$ by using \emph{cublasDmatinvBatched}. Using \emph{thrust::reduce} to calculate $\sum_{i=1}^V \beta_i$
        \item  $K$ is normally distributed as shown in equation \ref{eq35}. Samples from this distribution is taken by following the method discussed in section 3
    \end{enumerate}
    
    \item \textbf{Sample} $\mathbf{\rho}$
    \begin{enumerate}[(i)]
        \item This sampler has been discussed in section 3.
    \end{enumerate}
    
    \item \textbf{Sample} $\mathbf{\beta_i}$
    \begin{enumerate}[(i)]
        \item $\Lambda$ is copied to a vector $n_V$ number of times, the length of this vector is $(N-1)(N-1)n_V$. This is done so that $\Lambda_{\beta_i}$ in equation \ref{eq40} can be calculated by using $cublasDgemmBatched$
        \item $\Lambda_{\beta_i}^{-1}$ is calculated by \emph{cublasDmatinvBatched}. $\Lambda K$ is calculated by \emph{cublasDgemm}
        \item Equation \ref{eq40} is calculated by \emph{cublasDgemmBatched}
        \item All the $\beta_i$s are sampled in parallel by following the method to sample from a normal distribution discussed in the previous section. The \emph{batch size} should be $n_V$ here, because there are $n_V=V$ number of $\beta_i$s
    \end{enumerate}
\end{enumerate}

\subsection{Computing the sub-populations by using expectation maximization}

We use expectation maximization \cite{26} to find out the point estimates of the unknown parameters. The unknown parameters are $\Lambda$, $K$ and $\rho$. The update equations for our parallelizable model are shown below. Here, we have assumed the same data modeling as shown in equations \ref{eq3}, \ref{eq4} and prior assignments shown in equations \ref{eq5}, \ref{eq6}, \ref{eq7}.
Let $\Lambda^{(n)}$, $K^{(n)}$ and $\rho^{(n)}$ be the estimates at $n^{th}$ iteration. Here, $i$ ranges from 1 through $V$. 

By using expectation maximization, the update equations for the unknown parameters are:
\begin{equation}\label{eq45}
   \rho^{(n+1)}=\frac{V}{\sum_{i=1}^VS_i^{(n)}}
\end{equation}

\begin{equation}\label{eq46}
   K^{(n+1)}=\frac{\sum_{i=1}^VM_i^{(n)}}{V}
\end{equation}

\begin{equation}\label{eq47}
\begin{split}
    \Big\{\Lambda^{(n+1)}\Big\}^{-1}=\frac{1}{V}\sum_{i=1}^V\Big(&M_i^{(n)}M_i^{(n)T}+\Sigma_i^{(n)}\Big)
    -K^{(n+1)}K^{(n+1)T}
\end{split}
\end{equation}

Where, $\Sigma_i^{(n)}$, $M_i^{(n)}$ and $S_i^{(n)}$ respectively are

\begin{equation}\label{eq42}
    \Sigma_i^{(n)}=\Big[\Lambda^{(n)}+\rho^{(n)}D_i D_i^T\Big]^{-1}
\end{equation}

\begin{equation}\label{eq43}
    M_i^{(n)}=\Sigma_i^{(n)}\Big[\Lambda^{(n)}K^{(n)}+\rho^{(n)}D_i(r_i-\mu_i)\Big]
\end{equation}

\begin{equation}\label{eq44}
\begin{split}
        S_i^{(n)}=\{(r_i-&\mu_i)^2-2(r_i-\mu_i)D_i^TM_i^{(n)}+D_i^T(M_i^{(n)}M_i^{(n)T}+\Sigma_i^{(n)})D_i\}
\end{split}
\end{equation}
Figure \ref{em} explains the usage of expectation maximization to update the unknown parameters.
The procedure to compute the unknown parameters by using expectation maximization algorithm is as follows:
\begin{enumerate}[Step 1.]
    \item Compute all the $\Sigma_i$s in parallel 
    \item Compute all the $M_i$s in parallel 
    \item Use the $\Sigma_i$s and $M_i$s computed above and calculate all the $S_i$s in parallel
    \item Use the $S_i$ computed above and calculate $\rho$ as shown in equation \ref{eq45}
    \item Use the $M_i$ and compute $K$ as shown in equation \ref{eq46}
    \item Use the $M_i$, $\Sigma_i$ and $K$ computed above and calculate $\Lambda$ as shown in equation \ref{eq47}
\end{enumerate}
Now we shall calculate the point estimate of the unknown parameters on a GPU. We shall show that every computation is compatible with the SIMD architecture of a GPU. The steps are as follows.
\begin{enumerate}
    \item \textbf{Calculating $\mathbf{\Sigma_i^{(n)}}$}
    \begin{enumerate}[(i)]
        \item $\Lambda$ is copied $V$ number of times to a vector of length $(N-1)(N-1)V$, so that \emph{cublasDgemmBatched} followed by \emph{cublasDmatinvBatched} can be used to calculate all the $\Sigma_i^{(n)}$s in parallel on a GPU
    \end{enumerate}
    
    \item \textbf{Calculating $\mathbf{M_i^{(n)}}$}
    \begin{enumerate}[(i)]
        \item $\Lambda K$ is calculated by $cublasDgemm$
        \item $M_{i}^{(n)}$ is calculated by using \emph{cublasDgemmBatched}
    \end{enumerate}
    
    \item \textbf{Calculating $\mathbf{S_i^{(n)}}$}
    \begin{enumerate}[(i)]
        \item $(M_i^{(n)}M_i^{(n)T}+\Sigma_i^{(n)})$ is calculated by using \emph{cublasDgemmBatched}, the result is stored in a vector of length $(N-1)(N-1)V$
        \item $D_i^T(M_i^{(n)}M_i^{(n)T}+\Sigma_i^{(n)})$ is calculated by using \emph{cublasDgemmBatched}, the result is stored in a vector of length $(N-1)V$
        \item $D_i^T(M_i^{(n)}M_i^{(n)T}+\Sigma_i^{(n)})D_i$ is calculated by using \emph{cublasDgemmBatched}, the result is stored in a vector of length $V$
        \item $D_i^T(M_i^{(n)})$ is calculated by using \emph{cublasDgemmBatched}, the result is stored in a vector of length $V$
        \item $S_i^{(n)}$ is stored in a vector of length $V$
    \end{enumerate}
    \item To calculate $\mathbf{\rho^{(n+1)}}$, $\sum_{i=1}^VS_i^{(n)}$ is computed by $thrust::reduce$, then $\rho^{(n+1)}$ is calculated as shown in equation \ref{eq45}
    \item $\mathbf{\sum_{i=1}^V M_i^{(n)}}$ is computed by $thrust::reduce$, then $K^{(n+1)}$ is calculated as shown in equation \ref{eq46}
    
    \item \textbf{Calculating $\mathbf{\Big\{\Lambda^{(n+1)}\Big\}^{-1}}$}
    \begin{enumerate}[(i)]
        \item $\sum_{i=1}^V\Big(M_i^{(n)}M_i^{(n)T}+\Sigma_i^{(n)}\Big)$ is calculated by \newline \emph{thrust::reduce}
        \item $K^{(n+1)}K^{(n+1)}T$ is calculated by $cublasDgemm$
        \item $\Big\{\Lambda^{(n+1)}\Big\}^{-1}$ is calculated as shown in equation \ref{eq47}
    \end{enumerate}
\end{enumerate}

\subsection{Comparison of the methods}
The variational Bayes algorithm gives us the posterior distribution of the unknown parameters, and point estimates can be inferred from the posterior distribution. The expectation maximization gives us point estimates. The posterior distribution can be used to judge reliability of the results, because if the distributions are too spread out then the estimates can be considered unreliable. This gives the variational Bayes approach an advantage over expectation maximization. The prior knowledge about the weights can also be included in the prior distributions in Bayesian approach.\par
\begin{figure}
\centering
\begin{subfigure}{0.4\textwidth}
    \includegraphics[width=\textwidth]{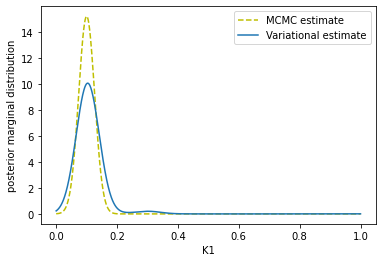}
    \caption{marginal of $K1$}
    \label{fig:first}
\end{subfigure}
\hfill
\begin{subfigure}{0.4\textwidth}
    \includegraphics[width=\textwidth]{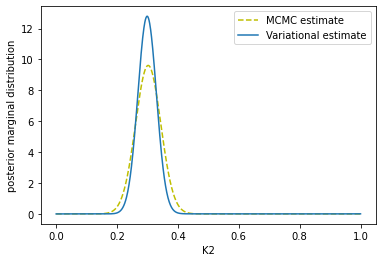}
    \caption{marginal of $K2$}
    \label{fig:second}
\end{subfigure}
\hfill
\begin{subfigure}{0.4\textwidth}
    \includegraphics[width=\textwidth]{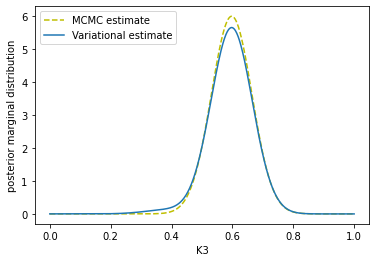}
    \caption{marginal of $K3$}
    \label{fig:third}
\end{subfigure}
\hfill
\begin{subfigure}{0.4\textwidth}
    \includegraphics[width=\textwidth]{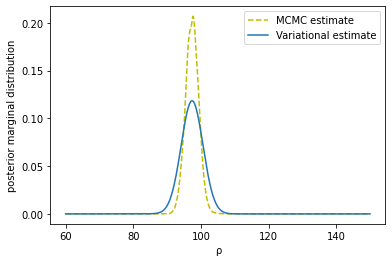}
    \caption{marginal of $\rho$}
    \label{fig:third}
\end{subfigure}        
\caption{Marginals of $K$ and $\rho$ computed from synthetic data.}
\label{fig4}
\end{figure}
\begin{figure}[htbp]
    \centering

   \begin{tikzpicture}
  \node (img)  {\includegraphics[width=7.9cm, height=4cm]{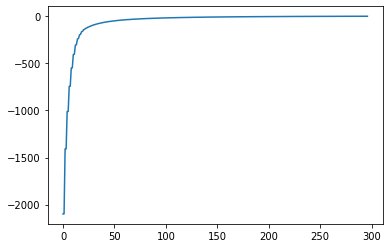}};
  \node[ node distance=0cm, rotate=90, anchor=center,yshift=4.0cm,xshift=0cm,font=\color{black}] {log lower bound};
  \node[node distance=0cm, rotate=0, anchor=center,yshift=-2.3cm,font=\color{black}] {iterations};
\end{tikzpicture}

    \caption{Increase of the log of lower bound with iterations of the variational Bayes for synthetic data}
    \label{figLB syn}
\end{figure}

In the synthetic data experiment, we ran the algorithms until the unknown parameters converged. Generally, the MCMC algorithm needs many iterations to converge compared to variational Bayes. We observed the same in our case. The variational algorithm and the expectation maximization algorithm converged in 100 iterations, and the MCMC algorithm converged in 8000 iteration. Kernel density estimation was done by using a multivariate Gaussian kernel \cite{23}. Figure \ref{fig4} shows the marginal distribution of $K$ and $\rho$ obtained from the variational Bayes and MCMC. The modes of $K$ obtained from the variational Bayes algorithm and Gibbs sampling are $(0.1031,\;0.2992,\;0.5977)^T$ and $(0.1011,\;0.3003,\;0.5986)^T$, these estimates are very close to the actual values of $K$ i.e. $(0.1,\;0.3,\;0.6)^T$. The $K$ estimates from the expectation maximization algorithm is $(0.1013,\;0.3013)^T$ and 99.312 is the estimate for $\rho$. All these estimates are very close to the actual value $(0.1,\; 0.3)^T$, thus showing the correctness of our model. $K3$ is estimated by $K3=1-K1-K2$. Figure \ref{fig4} shows how closely the posterior marginal distribution matches with both the methods. Figure \ref{figLB syn} shows the increase of the log of lower bound with iterations of the variational Bayes for synthetic data. The log lower bound stops changing after 100 iterations, thereby indicating convergence. Random restarts with different values of $K_{0}$ were attempted, the estimates always converged to the correct values of $K$ and $\rho$.

To check the correctness of our methods, we randomly generated $K$ vectors and calculated the Euclidean distance of the estimates from the actual value of $K$ for 5 different values of $K$. These $K$ are the first two elements from a Dirichlet distribution with parameter vector $(1\;1\;1)^T$. The average Euclidean distance error for the variational method, expectation maximization and the MCMC method are 0.009409, 0.0092103, 0.009881 showing that all the methods gave results close to the actual values with similar accuracy. Table 1 shows the results from variational Bayes, expectation maximization and MCMC algorithm for synthetic data for different values of $K$. The results from our model closely matches the actual value of $K$.\par

\begin{table}[htbp]

\caption{Table showing the results from variational Bayes, expectation maximization, and MCMC algorithm for synthetic data}
\label{tab:freq1}
\begin{tabular}{llll}
 \toprule
Actual value  & Variational Bayes & Expectation Maximization & Gibbs sampling  \\ \midrule
  $(0.1,\; 0.3)^T$ &  $(0.1031,\; 0.2992)^T$ &  $(0.1013,\; 0.3013)^T$ &  $(0.1011,\; 0.3003)^T$ \\ 
$(0.13,\; 0.25)^T$ &  $(0.1218,\; 0.2351)^T$ &  $(0.1218,\; 0.2341)^T$ &  $(0.1234,\; 0.2311)^T$ \\ 
$(0.39,\; 0.54)^T$ &  $(0.3814,\; 0.5342)^T$ &  $(0.3812,\; 0.5341)^T$ &  $(0.3829,\; 0.5312)^T$ \\ 
$(0.25,\; 0.16)^T$ &  $(0.2469,\; 0.1961)^T$ &  $(0.2468,\; 0.1939)^T$ &  $(0.2441,\; 0.1948)^T$ \\ 
$(0.18,\; 0.29)^T$ &  $(0.1791,\; 0.2678)^T$ &  $(0.1811,\; 0.2671)^T$ &  $(0.1822,\; 0.2643)^T$ \\ 
\bottomrule 
\end{tabular}


\end{table} 

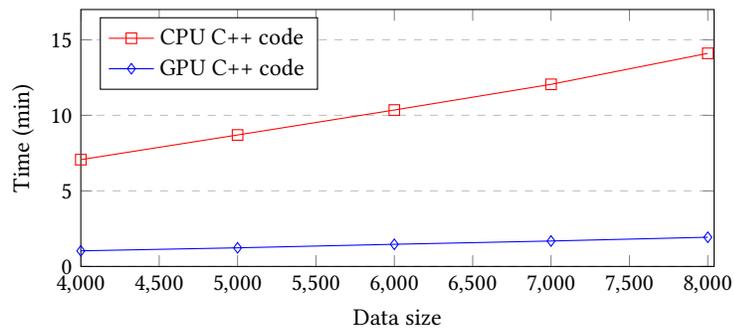
\begin{figure}[htbp]
\begin{tikzpicture}
\begin{axis}[
    xlabel={Data size},
    ylabel={Time (min)},
    xmin=4000, xmax=8040,
    ymin=0, ymax=17,
    legend pos=north west,
    ymajorgrids=true,
    grid style=dashed,
    ]

    \addplot[
    color=red,
    mark=square,
    ]
    coordinates {
    (4000,424/60)(5000,522/60)(6000,621/60)(7000,723/60)(8000,846/60)
    };

    \addplot[
    color=blue,
    mark=diamond,
    ]
    coordinates {
    (4000,62/60)(5000,74/60)(6000,88/60)(7000,101/60)(8000,116/60)
    };
    \addlegendentry{CPU C++ code}
    \addlegendentry{GPU C++ code}

\end{axis}
\end{tikzpicture}
\caption{Variational Bayes CPU and GPU run time for 4000 iterations}
\label{fig5}

\end{figure}

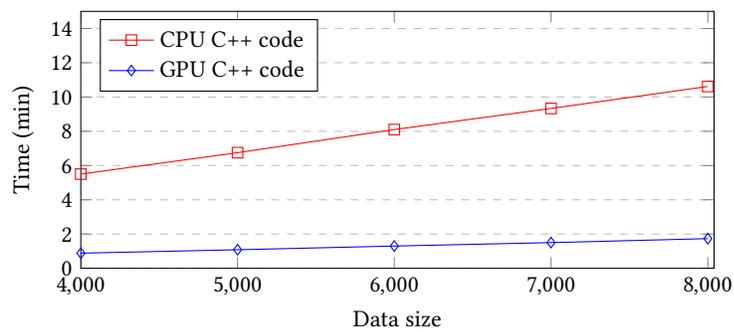
\begin{figure}[htbp]
\begin{tikzpicture}
\begin{axis}[
    xlabel={Data size},
    ylabel={Time (min)},
    xmin=4000, xmax=8040,
    ymin=0, ymax=15,
    xtick={4000,5000,6000,7000,8000},
    ytick={0,2,4,6,8,10,12,14,16},
    legend pos=north west,
    ymajorgrids=true,
    grid style=dashed,
    ]

    \addplot[
    color=red,
    mark=square,
    ]
    coordinates {
    (4000,330/60)(5000,405/60)(6000,486/60)(7000,560/60)(8000,637/60)
    };
    \addlegendentry{CPU C++ code}

    \addplot[
    color=blue,
    mark=diamond,
    ]
    coordinates {
    (4000,53/60)(5000,65/60)(6000,78/60)(7000,90/60)(8000,104/60)
    };
    \addlegendentry{GPU C++ code}
    
\end{axis}
\end{tikzpicture}
\caption{Expectation Maximization CPU and GPU run time for 4000 iterations}
\label{fig6}
\end{figure}

\begin{figure}[htbp]
\begin{tikzpicture}
\begin{axis}[
    xlabel={Data size},
    ylabel={Time (min)},
    xmin=4000, xmax=8040,
    ymin=0, ymax=46.5,
    xtick={4000,5000,6000,7000,8000},
    ytick={0,5,10,15,20,25,30,35,40},
    legend pos=north west,
    ymajorgrids=true,
    grid style=dashed,
    ]

    \addplot[
    color=red,
    mark=square,
    ]
    coordinates {
    (4000,1202/60)(5000,1490/60)(6000,1785/60)(7000,2104/60)(8000,2391/60)
    };
    \addlegendentry{CPU C++ code}

    \addplot[
    color=blue,
    mark=diamond,
    ]
    coordinates {
    (4000,201/60)(5000,246/60)(6000,297/60)(7000,341/60)(8000,389/60)
    };
    \addlegendentry{GPU C++ code}

\end{axis}
\end{tikzpicture}
\caption{CPU and GPU run time of MCMC for 10,000 Monte Carlo iterations}
\label{fig7}
\end{figure}
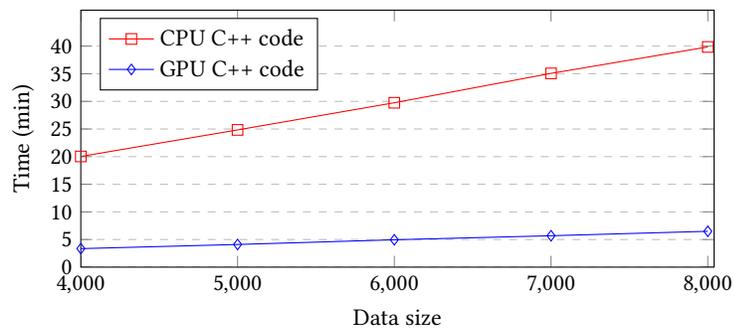

\noindent We used a Nvidia GeForce GTX 1050 with 4GB RAM GPU to run our GPU codes. To compare the run time, we ran our algorithms on a laptop GPU and CPU. We wrote the GPU codes on CUDA C++, and the CPU codes were run on Intel Core i5 8300H CPU. To compare the run times we changed the data size from 4000 to 8000. The run time comparison for the three algorithms is shown in figures \ref{fig5}, \ref{fig6} and \ref{fig7} respectively. These figures show that the CPU run time increases with data size, but the GPU run time is not much affected by this increase in data size. Figure \ref{fig5} and \ref{fig6} also show that the GPU run time for the variational Bayes and expectation maximization algorithm is almost the same. The GPU run time of the variational Bayes and expectation maximization algorithm is less compared to the GPU run time of Gibbs sampling.

The variational Bayes model is the best because it converges faster than the MCMC model and gives us the posterior distribution of the unknown distributions; point estimates can be made from the distributions. Expectation maximization gives us only the point estimates of the unknown parameters, and these point estimates have no information about the reliability of the results. In variational Bayes, we can judge the reliability of the results; if the distribution is too spread, the estimates are considered unreliable, and we could consider collecting more data. Therefore the variational Bayes method is superior to the MCMC and expectation maximization method. 
\subsection{Demonstration using experimental data}
To test our developed model we used experimental data from real fibroblast tissues \cite{8}. 
Authors in \cite{g}, \cite{8}, \cite{h} have used the same dataset to compute the weight of each network. The dataset has two groups 1 and 2, depending on the drugs to which the tissues are subjected to. The dataset has 56 observed gene expression measurements and expression profiles associated with each measurement, so $V=56$. The authors in \cite{8} assumed three networks, i.e., $N=3$. Network 1 does not have any stuck at faults. Network 2 has a stuck-at one fault at ERK1/2 and network 3 has a stuck-at one fault at SRF-ELK1 and SRF-ELK4.\par

\begin{figure}[htbp]
    \centering
   \begin{tikzpicture}
  \node (img)  {\includegraphics[width=7.9cm, height=4cm]{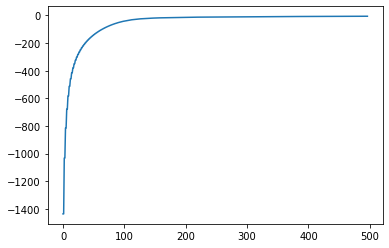}};
  \node[ node distance=0cm, rotate=90, anchor=center,yshift=4.0cm,xshift=0cm,font=\color{black}] {log lower bound};
  \node[node distance=0cm, rotate=0, anchor=center,yshift=-2.3cm,font=\color{black}] {iterations};
\end{tikzpicture}

    \caption{Increase of the log of lower bound with iterations of the variational Bayes for real experimental data}
    \label{figLBex}
\end{figure}

\begin{figure}
\centering
\begin{subfigure}{0.4\textwidth}
    \includegraphics[width=\textwidth]{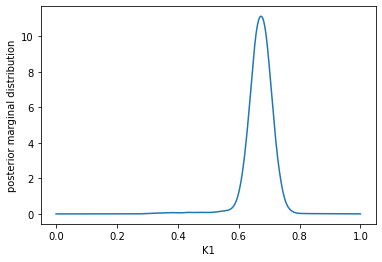}
    \caption{marginal of $K1$}
    \label{fig:first}
\end{subfigure}
\hfill
\begin{subfigure}{0.4\textwidth}
    \includegraphics[width=\textwidth]{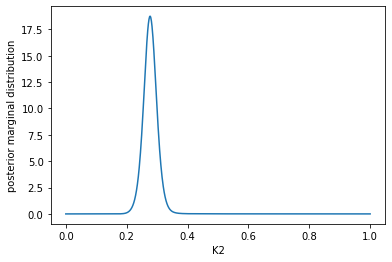}
    \caption{marginal of $K2$}
    \label{fig:second}
\end{subfigure}
\hfill
\begin{subfigure}{0.4\textwidth}
    \includegraphics[width=\textwidth]{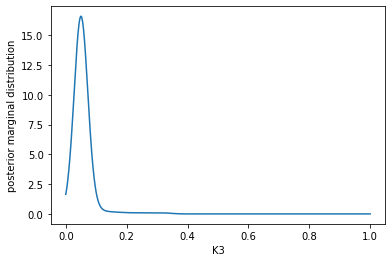}
    \caption{marginal of $K3$}
    \label{fig:third}
\end{subfigure}
\hfill
\begin{subfigure}{0.4\textwidth}
    \includegraphics[width=\textwidth]{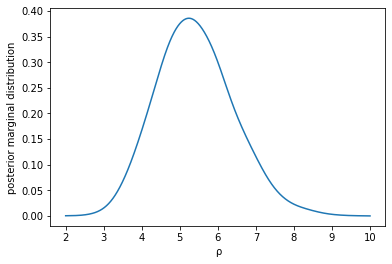}
    \caption{marginal of $\rho$}
    \label{fig:third}
\end{subfigure}        
\caption{Marginals of $K$ and $\rho$ computed from experimental data.}
\label{fig8}
\end{figure}

The marginal distribution of the $K$s computed by the variational Bayes algorithm on a GPU are shown in figure \ref{fig8}. The modes of the marginals of $K$'s are $(0.6676,\;0.2782,\;0.0542)^T$. The mode of $\rho$ computed by variational Bayes is 5.26.  The results from MCMC are not shown because it could produce good effective sample size. Figure \ref{figLBex} shows the increase in the log of lower bound with iterations of the variational Bayes algorithm for real experimental data. The log lower bound stops updating after 200 iterations, therefore indicating convergence. The estimates of $K$'s from the expectation maximization algorithm is $(0.6750,\;0.2764,\;0.0486)^T$ and the estimate for $\rho$ is 5.57. The estimates are very close to results obtained in \cite{8} i.e., $(0.6716,\;0.2740,\;0.0544)^T$ from the variational Bayes algorithm and $(0.6764,\;0.2745,\;0.0490)^T$ from the expectation maximization algorithm. The results agree with our previous results obtained in \cite{h} i.e., $(0.6161,\;0.3236,\;0.0603)^T$. This shows that the first subpopulation is the dominant one with a weightage of 61.6\%. 

\begin{table}
  \caption{Table showing the results from real experimental data \cite{8} and computation time of current methods. Our method is parallelizable and converges faster than \cite{h}. Here $n_V$ is the number of gene expression data, $m$ is the number of Markov chain Monte Carlo iteration and $i$ is the number of variational Bayes iterations, where $i<n_V$ and $m<n_V$}
  \label{tab:freq}
  \begin{tabular}{llll}
    \toprule
    Method & Results & Complexity & Iterations to\\& & & converge\\
    \midrule
    MCMC\cite{g} &  $(0.6453,\; 0.2255,\; 0.1292)^T$ & $n_V$ & 100K \\
    Variational methods\cite{8} &  $( 0.6716,\; 0.2740,\;0.0544 )^T$ & $n_V$ & 200 \\
    Parallelizable MCMC\cite{h} &  $(0.6161,\;0.3236,\;0.0603)^T$ & $m$ & 100K\\
    \textbf{Proposed model} &  \boldmath{$(0.6676,\;0.2782,\;0.0542)^T$} & \textbf{$i$} & \textbf{200} \\
  \bottomrule
\end{tabular}
\end{table}

\section{Conclusion and future work}
In this paper we presented a new efficient parallelizable model to compute the compositional breakup of cancerous tissue. The model uses a collection of Boolean networks to model the subpopulations. Using synthetic and experimental data, we have demonstrated the usage of this model. This paper presented a new parallelizable model with reduced complexity and thereby reducing computation time. The variational Bayes algorithm was used to estimate the subpopulation breakup. The effectiveness of the variational Bayes algorithm was confirmed by comparing the estimates with expectation maximization and MCMC method on a GPU. We have compared our results, complexity, and iterations required to converge with other models in table 2. The model described in \cite{g}, and \cite{8} are not parallelizable, so their complexity increases with the size of input data; as a result, the computation time increases. The model discussed in \cite{h} is compatible with a GPU. The complexity of the model discussed in \cite{h} does not depend on the size of input data; as a result, the computation time does not depend much on the input data size, but it is computationally intensive. The model in \cite{h} uses the MCMC algorithm, so it takes a lot of iterations to converge. Whereas our new model is parallelizable, so the complexity does not change with the size of the data and can achieve the same results in fewer iterations than \cite{h}. We also demonstrated the usage of variational Bayes method to estimate the unknown parameters of a hierarchical model on a GPU. This helped us to reduce the computation time.\par
In this paper, we accelerated variational Bayes and expectation maximization for GPU. Such acceleration can also be done with distributed systems, where the workload can be divided among many compute nodes, and the final result can be gathered from the master node.

\bibliographystyle{ACM-Reference-Format}

\appendix

\end{document}